\documentclass[a4paper,fleqn]{cas-dc}

\usepackage{graphicx} 
\usepackage{amsmath} 
\usepackage{cite} 
\usepackage{siunitx} 

\def\tsc#1{\csdef{#1}{\textsc{\lowercase{#1}}\xspace}}
\tsc{WGM}
\tsc{QE}

\begin{document}
\let\WriteBookmarks\relax
\def\floatpagepagefraction{1}
\def\textpagefraction{.001}

\shorttitle{Towards Sustainable Horizons: A Comprehensive Blueprint for Mars Colonization}    

\shortauthors{Florian Neukart}  

\title [mode = title]{Towards Sustainable Horizons: A Comprehensive Blueprint for Mars Colonization}  



%
\author[1,2]{Florian Neukart}[type=author,
      style=chinese,
      auid=000,
      bioid=1,
      orcid=0000-0002-2562-1618]

\affiliation[1]{organization={Leiden Institute of Advanced Computer Science},
            addressline={Snellius Gebouw, Niels Bohrweg 1}, 
            city={Leiden},
            postcode={2333 CA}, 
            state={South Holland},
            country={Netherlands}}

\affiliation[2]{organization={Terra Quantum AG},
            addressline={Kornhausstrasse 25}, 
            city={St. Gallen},
            postcode={9000}, 
            state={St. Gallen},
            country={Switzerland}}

\begin{abstract}
Establishing a human colony on Mars is one of the most ambitious endeavors of our time. This paper provides a comprehensive assessment of the challenges and solutions related to Mars colonization, emphasizing sustainability, efficiency, and the well-being of colonists. We begin by analyzing the Martian environment, focusing on challenges such as radiation, dust storms, temperature variations, and low atmospheric pressure. The discourse then transitions into technological solutions, exploring innovations in infrastructure, energy production, transportation, and life support systems. Special attention is paid to harnessing in-situ resources and recent advancements like Martian concrete, aeroponics, and algae bioreactors. The human dimension is addressed, from the psychological implications of prolonged isolation to physiological considerations in reduced gravity. Economic considerations encapsulate the cost-benefit analysis of in-situ resource utilization versus Earth transport and the potential incentives for private sector investment. The paper culminates in recommendations for future research, highlighting areas pivotal for refining the blueprint of Mars colonization. This work serves as a foundational guide for researchers, policymakers, and visionaries aiming to make humanity's interplanetary future a reality.
\end{abstract}

\begin{highlights}
\item Detailed analysis of Mars' environmental challenges, from dust storms to radiation.
\item Exploration of sustainable infrastructure solutions, including Martian concrete and underground cities.
\item Examination of diverse energy production methods suitable for Mars, from solar to nuclear.
\item Emphasis on innovative transportation methods, including solar sail-powered cargo ships and specialized rovers.
\item Comprehensive look at life support systems, emphasizing water harvesting, aeroponics, and bio-reactors.
\item Discussion on the psychological and physiological challenges faced by astronauts during extended Mars missions.
\item Economic considerations touching on the costs of utilizing in-situ resources versus transport from Earth and potential private sector incentives.
\end{highlights}


\maketitle

\sloppy
\section{Introduction}

The quest to transcend our planetary boundaries and establish a presence on another celestial body is one of the most ambitious endeavors undertaken by humanity \cite{Zubrin1996, Fogg1995}. Mars, with its relative proximity and some similarities to Earth, emerges as the prime candidate for colonization \cite{Cockell2001}. Establishing a sustainable human presence on the Red Planet is a testament to human engineering and determination and may also act as a potential safeguard against global threats to humanity \cite{Baum2008}.

\subsection{Need for Colonization}

The rationale behind colonizing Mars is multifaceted \cite{Zubrin1996}:
\begin{itemize}
\item \textbf{Planetary Backup}: Earth, though currently our only home, is susceptible to global catastrophes \cite{Baum2008}. These can range from natural disasters, such as asteroid impacts or supervolcano eruptions, to anthropogenic threats, like nuclear warfare or runaway climate change \cite{Bostrom2002}.
\item \textbf{Scientific Exploration}: Mars offers a treasure trove of scientific data \cite{Carr2006}. Unraveling its geological and climatic history can provide insights into planetary evolution, the prevalence of life, and the conditions required for its existence \cite{Grotzinger2014}.
\item \textbf{Technological Advancement}: The challenges of Mars colonization drive technological and engineering advancements that can have terrestrial applications \cite{Zubrin2011}, potentially leading to breakthroughs in energy storage, life support systems, and robotics \cite{Sherwood2009}.
\item \textbf{Inspiration}: Beyond practical reasons, venturing into the unknown has always sparked human imagination and served as a testament to our indomitable spirit \cite{McKay2004}.
\end{itemize}

\subsection{Current Challenges}

While the allure of Mars is compelling, the challenges are daunting \cite{Zubrin2011}:
\begin{itemize}
\item The \textit{vast distance} from Earth means that resupply missions would be infrequent and communication would have non-negligible latency \cite{Ley2016}.
\item Mars' \textit{thin atmosphere} and the absence of a magnetosphere expose settlers to heightened radiation levels \cite{Hassler2014}.
\item The \textit{low temperatures} and reduced atmospheric pressure require specialized habitats and life support systems \cite{Sherwood2009}.
\item The \textit{gravity on Mars}, about 38\% that of Earth's, may have long-term physiological effects on settlers \cite{Clément2015}.
\end{itemize}

\subsection{Significance of Sustainable and Cost-Effective Methods}

Given the prolonged mission durations, immense distances, and resource constraints, traditional "Earth-reliant" approaches to space exploration are not feasible for Martian colonization \cite{Moses2012}. Sustainability is paramount:
\begin{itemize}
\item \textbf{Economic Feasibility}: Repeatedly shipping materials and resources from Earth to Mars is prohibitively expensive \cite{Kleinhenz2018}. Utilizing in-situ resources alleviates this, making missions economically viable \cite{Zacny2015}.
\item \textbf{Reduced Dependence}: Due to the long communication and travel time between the planets, Martian settlements must aim for self-sufficiency regarding energy, food, and other essential resources \cite{Clément2015}.
\item \textbf{Environmental Impact}: Just as we grapple with environmental concerns on Earth, we must ensure that our Martian endeavors are conducted responsibly to prevent irreversible changes to the Martian environment \cite{Cockell2016}.
\end{itemize}
This publication describes the nuances of establishing a sustainable and efficient human colony on Mars, exploring state-of-the-art technologies and engineering solutions, and addressing the multidisciplinary challenges that arise \cite{Zubrin2011, McKay2004}.

\section{Background}
\label{sec:background}

Mars, often dubbed the `Red Planet', has been an object of human curiosity and scientific investigation for centuries. With advances in space technology over the latter half of the 20th century and into the 21st century, our understanding of Mars has transitioned from telescopic observations to detailed in-situ investigations by robotic emissaries \cite{Wall2011, Carr2006}.

\subsection{Early Observations and Speculations}
\label{subsec:early_observations}

Mars has been known since ancient times due to its visible reddish appearance in the night sky. The planet's distinctive color is due to iron oxide (rust) on its surface \cite{Bell2004}.

Observations in the late 19th century by astronomers such as \textit{Giovanni Schiaparelli} and later \textit{Percival Lowell} led to the popular, albeit incorrect, notion of `canals' on Mars, possibly signifying intelligent life forms. Lowell's extensive publications and lectures on his observations and theories about the Martian canals fed public interest and set the stage for subsequent scientific investigation \cite{Sheehan1996, Dick1988}.

\subsection{Robotic Exploration}
\label{subsec:robotic_exploration}

\subsubsection{Early Probes and Flybys}
\label{subsubsec:early_probes}

The space age initiated the era of robotic Mars exploration. The first attempts by the Soviet Union's \textit{Mars} program in the early 1960s faced challenges, with several missions failing either during launch, en-route, or upon arrival \cite{Harvey1996}.

The USA's \textit{Mariner} program was the first to successfully send back data from Mars \cite{Kieffer1992}. \textit{Mariner 4}, in 1965, performed the first successful flyby, revealing a barren landscape, contrary to previous speculations of a wetter Mars.

\subsubsection{Landers and the Quest for Life}
\label{subsubsec:landers}

The 1970s saw the ambitious \textit{Viking} program by NASA, aiming to land on and study the Martian surface directly. \textit{Viking 1} and \textit{Viking 2} landers undertook experiments to detect microbial life. Although the results from the labeled release experiment initially indicated possible metabolic activity, subsequent analysis suggested that the reactions were more likely due to inorganic processes. The quest for life thus remained inconclusive \cite{Klein2015, Stoker2010}.

\subsubsection{Rovers: Mobile Geology Labs}
\label{subsubsec:rovers}

In the late 1990s and into the 2000s, rovers became central to Mars exploration. \textit{Pathfinder's} Sojourner was the first, followed by the twin rovers \textit{Spirit} and \textit{Opportunity}. These missions revolutionized our understanding of Martian geology and climate history \cite{Arvidson2011}. Notably, \textit{Opportunity} discovered evidence of ancient liquid water flows on the Martian surface \cite{Squyres2004}.

The car-sized \textit{Curiosity} rover, part of NASA's \textit{Mars Science Laboratory} mission, landed in 2012. Equipped with a suite of sophisticated instruments, \textit{Curiosity} identified an ancient lakebed and detected organic molecules, further intensifying the search for potential past life \cite{Grotzinger2014, Eigenbrode2018}.

\subsubsection{Orbiters: Eyes in the Martian Sky}

 Complementing the surface missions, a series of orbiters have provided crucial data about Mars' atmosphere, climate, and geology. Notable missions include NASA's \textit{Mars Reconnaissance Orbiter} (MRO) and ESA's \textit{Mars Express} \cite{McEwen2007, Chicarro2004}. MRO's HiRISE camera has captured high-resolution images, aiding in landing site selection and studying geological features, while its SHARAD instrument has probed subsurface layers.

\subsection{Recent Endeavors and International Collaboration}

 The 2020s saw a global convergence towards Mars exploration. NASA's \textit{Perseverance} rover and its innovative \textit{Ingenuity} helicopter, the UAE's \textit{Hope} orbiter, and China's \textit{Tianwen-1} mission, which includes both an orbiter and a rover, all reached Mars in 2021 \cite{Witze2021, Amos2021}. Collaborative efforts, both international and between governmental and private entities, are seen as the pathway forward.

\subsection{Prelude to Human Exploration}

 With the wealth of data from robotic missions, plans for human exploration and eventual colonization have gained momentum. Both NASA and ESA have outlined potential manned missions to Mars in their future roadmaps, viewing the Moon as a testing ground. Private entities, particularly SpaceX, envision large-scale colonization efforts \cite{Foust2019,Wall2020}.

 While the allure of Mars colonization is compelling, a careful synthesis of our learnings from robotic missions, rigorous research into life support and habitat systems, and international collaboration are vital to turn this vision into a reality \cite{Newman2015}.

\section{Martian Environment and Challenges}

\subsection{Radiation}

 One of the most pressing challenges for human exploration and potential colonization of Mars is the intense radiation from galactic cosmic rays (GCRs) and solar energetic particles (SEPs) \cite{Zeitlin2013}. Unlike Earth, Mars lacks a strong magnetosphere and a thick atmosphere, both of which significantly attenuate the incoming cosmic and solar radiation on our home planet.

\subsubsection{Sources of Radiation}

 Two primary sources of space radiation would affect Mars explorers:

\begin{itemize}
    \item \textbf{Galactic Cosmic Rays (GCRs):} These are high-energy protons and heavy ions that originate from sources outside our solar system, including supernovae. Their energy levels are so high that they can penetrate even the most advanced shielding technologies currently available \cite{Cucinotta2018}.
    \item \textbf{Solar Energetic Particles (SEPs):} These are sporadic bursts of radiation primarily composed of protons, emitted from the Sun during solar flares and coronal mass ejections \cite{Reames2013}.
\end{itemize}

\subsubsection{Effects on Human Health}

 Prolonged exposure to space radiation has several potential health effects \cite{Durante2008}:

\begin{itemize}
    \item Increased risk of cancer.
    \item Acute radiation sickness from intense radiation bursts, such as significant solar flares.
    \item Degenerative tissue effects, including cataract formation, cardiovascular diseases, and potential central nervous system damage.
    \item Possible infertility and hereditary effects.
\end{itemize}

\subsubsection{Quantifying Martian Radiation}

 Measurements by the \textit{Radiation Assessment Detector (RAD)} on the \textit{Curiosity} rover revealed that an astronaut would be exposed to a minimum of 0.66 sieverts during a round trip to Mars, with a stay on the surface \cite{Zeitlin2013}. This is more than three times the radiation limit recommended for astronauts during their entire career.

 Given Eq. \eqref{eq:radiation_dose}:

\begin{equation}
    D = \int_{0}^{T} R(t) dt
\label{eq:radiation_dose}
\end{equation}

where \(D\) is the cumulative radiation dose in sieverts (Sv), \(R(t)\) is the radiation rate as a function of time, and \(T\) is the total time of exposure, we can integrate over the duration of the Mars mission to determine the total dose an astronaut would receive.

\subsubsection{Mitigating Radiation Exposure}

 Several strategies are under consideration to mitigate the radiation risk:

\begin{itemize}
    \item \textbf{Advanced Shielding:} Using materials such as water, polyethylene, or novel materials like hydrogen-rich boron nitride nanotubes for spacecraft and habitats \cite{Townsend1992, Durante2011}.
    \item \textbf{Underground Habitats:} Constructing habitats below the Martian surface can use Mars' regolith as a natural radiation barrier \cite{Cucinotta2014}.
    \item \textbf{Magnetic Shields:} Generating a protective magnetosphere around the habitat or spacecraft requires significant energy \cite{Hassler2014, Durante2011}.
    \item \textbf{Pharmaceutical Countermeasures:} Drugs that can mitigate radiation damage or boost the body's natural repair mechanisms \cite{Durante2017}.
\end{itemize}

 Conclusively, while radiation is a significant challenge for Mars exploration and colonization, ongoing research aims to develop viable solutions to ensure the safety and health of future Martian explorers and settlers \cite{Cucinotta2014}.

\subsection{Dust Storms}
Mars is infamous for its planet-wide dust storms, which can last for weeks to months, enveloping the entire surface in a thick haze \cite{Smith2018}. While these storms are not accompanied by the high winds seen in terrestrial storms due to the thin Martian atmosphere \cite{Newman2017}, they pose unique challenges for human exploration and colonization.

\subsubsection{Characteristics of Martian Dust Storms}
\begin{itemize}
    \item \textbf{Frequency:} Localized dust storms frequently occur on Mars, but approximately once every three Martian years (or roughly 5.5 Earth years), the conditions allow these local storms to expand and merge, covering the entire planet \cite{Smith2018}.
    \item \textbf{Composition:} Martian dust particles are extremely fine, comparable to talcum powder, and are composed mainly of iron oxides, giving Mars its characteristic red color \cite{Bridges2012}. 
    \item \textbf{Electrostatic Nature:} Due to the continuous bombardment by ultraviolet (UV) radiation, these dust particles can become electrostatically charged, causing them to cling to surfaces \cite{Esposito2006}.
\end{itemize}

\subsubsection{Challenges Posed by Dust Storms}
\begin{itemize}
    \item \textbf{Solar Power Generation:} One of the immediate challenges of dust storms is the significant reduction in sunlight, which can drop to less than 1\% of normal levels during a global dust storm \cite{Lemmon2015}. This poses a threat to missions that rely on solar power. For instance, the Opportunity rover's mission was terminated after it lost power during a massive dust storm in 2018 \cite{Bell2019}.
    \item \textbf{Thermal Management:} Dust accumulation on equipment and habitats can disrupt thermal control, either by insulating equipment or by blocking radiators, leading to overheating \cite{Martinez2019}.
    \item \textbf{Equipment Wear and Tear:} The fine, abrasive nature of Martian dust can lead to wear and tear on equipment, especially moving parts like joints and wheels on rovers or seals on habitats \cite{Bridges2012}.
    \item \textbf{Human Health:} The dust's fine nature poses a potential respiratory hazard for astronauts. If it's tracked inside habitats, it could be inhaled and cause health issues \cite{James2013}. Additionally, the iron oxides in the dust might pose chemical risks if ingested or inhaled in significant amounts \cite{Drake2018}.
    \item \textbf{Visibility and Navigation:} Dust storms can hamper visibility, making navigation and scientific observations challenging \cite{Smith2018}.
\end{itemize}

\subsubsection{Mitigation Strategies}
\begin{description}
    \item[Alternative Power Sources:] Exploratory missions and habitats can integrate backup power sources, such as nuclear power or advanced battery storage, to remain operational during prolonged dust storms \cite{Drake2018, Lemmon2015}.
    \item[Protective Coatings:] Equipment can be coated with materials that repel electrostatically charged dust, reducing accumulation \cite{Esposito2006}.
    \item[Airlocks and Filtration:] Advanced airlock systems and filtration units can reduce the amount of dust brought into Martian habitats, protecting both equipment and astronauts \cite{James2013}.
    \item[Dust Monitoring and Forecasting:] Integrating meteorological equipment to monitor and forecast dust movements can allow missions to take preventive measures before storms hit their location \cite{Martinez2019}.
\end{description}

While Martian dust storms do not pose an immediate life-threatening danger, as Hollywood might suggest, they do present substantial operational challenges \cite{Lemmon2015, Bell2019}. Future Martian missions must be equipped to handle these challenges to ensure the safety and success of human exploration and colonization efforts.

\subsection{Temperature Variations}
Mars, often referred to as the "Red Planet," exhibits extreme temperature fluctuations, much more pronounced than those experienced on Earth \cite{Haberle1996, Jakosky2018}. This vast temperature range, combined with the thin atmosphere, has direct implications for human habitation and technological equipment functioning on the Martian surface.

\subsubsection{Characteristics of Martian Temperature}
\begin{itemize}
    \item \textbf{Daily Fluctuations:} Mars experiences significant diurnal temperature fluctuations, with temperatures at the equator ranging from a maximum of about 20°C (70°F) at noon to a minimum of -73°C (-100°F) at night \cite{Smith2004}.
    \item \textbf{Seasonal Variations:} Owing to its axial tilt of 25.2 degrees, almost similar to Earth's 23.5 degrees, Mars undergoes seasons \cite{Jakosky2018}. However, the lengths of the Martian seasons are about twice as long, given its 687-day year. Winters, especially in the polar regions, can see temperatures drop to -125°C (-195°F) \cite{Haberle1996}.
    \item \textbf{Atmospheric Influence:} Mars' thin atmosphere, composed mainly of carbon dioxide, does not trap heat effectively \cite{Forget1999}. This lack of a substantial greenhouse effect is a key reason for the planet's cold temperatures.
\end{itemize}

\subsubsection{Challenges Arising from Temperature Variations}
\begin{itemize}
    \item \textbf{Human Survival:} The extremely cold temperatures can be fatal to humans without appropriate protection \cite{Rummel2004}.
    \item \textbf{Equipment Functionality:} Electronic and mechanical equipment can malfunction or fail when exposed to such low temperatures \cite{Zacny2009}. Lubricants can freeze, metals can become brittle, and batteries can lose charge rapidly.
    \item \textbf{Thermal Expansion and Contraction:} The significant temperature fluctuations can cause materials to continuously expand and contract, leading to material fatigue and potential structural failures over time \cite{Ulrich2010}.
    \item \textbf{Energy Demands:} Maintaining a habitable temperature inside Martian bases would require a consistent and substantial energy source, especially during the long, cold nights and winters \cite{Hecht2009}.
\end{itemize}

\subsubsection{Mitigation Strategies}
\begin{itemize}
    \item \textbf{Insulated Habitats:} Designing habitats with advanced insulating materials can help maintain a stable internal temperature \cite{Hoffman2016}.
    \item \textbf{Subsurface Habitats:} Building habitats below the Martian surface can leverage the ground's natural insulation properties, leading to more stable internal temperatures \cite{Boston2004}.
    \item \textbf{Thermal Blankets and Heaters for Equipment:} Protecting equipment with thermal blankets and integrating heaters can prevent damage from cold temperatures \cite{Zacny2009}.
    \item \textbf{Utilizing Local Resources:} Ice deposits on Mars can be melted and used in heat exchange systems to help regulate temperatures inside habitats \cite{Hecht2009}.
\end{itemize}

Adapting to the extreme and fluctuating temperatures of Mars is paramount for the success of future manned missions and potential colonization efforts \cite{Hoffman2016}. Advanced technologies and innovative solutions will be essential to ensure astronauts' safety and equipment's durability in such a challenging environment.

\subsection{Low Atmospheric Pressure}
The atmospheric pressure on Mars is significantly lower than Earth's, averaging around 600 pascals (0.087 psi), which is less than 1\% of Earth's mean sea level pressure \cite{Smith2001, Haberle2017}. This meager atmospheric pressure, primarily composed of carbon dioxide (\(CO_2\)), presents unique challenges for human missions and colonization \cite{Zubrin1996}.

\subsubsection{Characteristics of Martian Atmospheric Pressure}
\begin{itemize}
    \item \textbf{Composition:} Mars' atmosphere is about 95\% \(CO_2\), with traces of nitrogen and argon \cite{Jakosky1992}. The absence of a substantial oxygen component starkly contrasts Earth's oxygen-rich atmosphere.
    \item \textbf{Altitude Variations:} The atmospheric pressure on Mars varies with altitude, with the highest pressures (up to 1,155 pascals) found in the depths of the Hellas Basin and the lowest pressures at the top of Olympus Mons \cite{Muhleman1991}.
    \item \textbf{Seasonal Variations:} Martian atmospheric pressure exhibits slight seasonal variations due to the sublimation and deposition of \(CO_2\) from and onto the polar ice caps \cite{Hess1979}.
\end{itemize}

\subsubsection{Challenges Arising from Low Atmospheric Pressure}
\begin{itemize}
    \item \textbf{Human Exposure:} Direct exposure to Mars' low-pressure atmosphere would be lethal to humans \cite{Webb1998}. In such conditions, bodily fluids would vaporize, leading to a condition called ebullism.
    \item \textbf{EVA Suit Design:} Designing space suits for extended extravehicular activity (EVA) on Mars is challenging due to balancing mobility and maintaining a higher internal pressure \cite{Newman2000}.
    \item \textbf{Aerodynamic Behavior:} The thin atmosphere affects the aerodynamics of vehicles, making landing and takeoff operations tricky. Traditional parachutes become less effective, demanding alternative or supplemental landing technologies \cite{Mitchell2016}.
    \item \textbf{Radiation:} The thin atmosphere provides limited shielding from harmful solar and cosmic radiation \cite{Hassler2014}.
\end{itemize}

\subsubsection{Mitigation Strategies}
\begin{itemize}
    \item \textbf{Pressurized Habitats:} Designing habitats that can withstand the internal-external pressure differential is crucial \cite{Carr2006}. These habitats would maintain an Earth-like atmosphere inside to support human life.
    \item \textbf{Advanced EVA Suits:]} New EVA suit designs could utilize mechanical counterpressure to apply direct pressure to the skin, enhancing mobility while maintaining safety \cite{Aitchison2011}.
    \item \textbf{Innovative Landing Systems:} Incorporating retro-rockets, inflatable heat shields, or sky cranes can assist in safe landings in the thin Martian atmosphere \cite{Munk2012}.
    \item \textbf{Local Resource Utilization:} Extracting \(O_2\) from local resources, such as water ice or the carbon dioxide atmosphere itself (via electrolysis or chemical processes), can help sustain human colonies and refuel rockets \cite{Mckay1992}.
\end{itemize}

Mars's uniquely low atmospheric pressure necessitates innovative engineering and biomedical solutions \cite{Zubrin2011}. Addressing these challenges is fundamental for ensuring the safety and success of future Martian exploratory and colonization efforts.

\section{Technological Solutions}
 As humanity contemplates the formidable task of establishing a presence on Mars, it becomes clear that existing technologies—while groundbreaking—are not sufficient. Adapting to Mars' unique challenges mandates fresh technological innovations, particularly those that leverage Mars' own resources and the unique aspects of its environment \cite{Zubrin2011, Carr2006}. This section describes the field of Martian-specific technology solutions, spanning infrastructure, resource utilization, transport, and life support.

\subsection{Infrastructure \& Habitat}
 One of the foremost challenges of Martian colonization lies in creating structures that shelter astronauts from the planet's harsh environment and promote sustained human habitation. Habitats on Mars must be robust against radiation, temperature fluctuations, and low atmospheric pressures, all while being feasible to construct with limited resources transported from Earth \cite{Hassler2014}. Hence, the ideal approach would lean heavily on in-situ resource utilization, transforming Martian raw materials into sturdy, long-lasting habitats \cite{Mckay1992}. This subsection explores the latest advancements and proposals, examining how we might construct our first homes on another planet.

\subsubsection{Martian Concrete}
 For sustainable human colonization on Mars, it is pivotal to utilize local resources, minimizing the need for resource-intensive shipments from Earth \cite{Mckay1992}. An essential development in this realm is the prospect of Martian concrete using native regolith and Martian-abundant materials \cite{Lin2016}.

\paragraph{\textbf{Characteristics of Martian Regolith}}
\begin{itemize}
    \item \textbf{Composition:} Martian regolith comprises finely milled rock, soil, and dust \cite{Carr2006}. Its iron oxide content imparts the planet's distinctive red color. This regolith spans a spectrum of particle sizes, from fine dust to small rocks.
    \item \textbf{Basaltic Nature:} A predominant component of Martian soil is weathered basalt, a volcanic rock common to the planet \cite{Carr2006}.
    \item \textbf{Water Content:} Recent revelations indicate that bound water exists in Mars' regolith, albeit in trace amounts \cite{Carr2006}.
\end{itemize}

\paragraph{\textbf{Formulation of Martian Concrete}}
\begin{itemize}
    \item \textbf{Binding Agent:} One innovative proposal involves using sulfur, abundant on Mars, as a binding agent. After heating, sulfur transitions to a liquid state and solidifies upon cooling, effectively binding the regolith \cite{Lin2016}.
    \item \textbf{Strength and Durability:} Preliminary Earth-based trials with Martian regolith simulants and sulfur have demonstrated a compressive strength on par with terrestrial concrete \cite{Lin2016}. It also encourages resistance to Martian-specific challenges, such as radiation and temperature swings.
\end{itemize}

\paragraph{\textbf{Advantages of Martian Concrete}}
\begin{itemize}
    \item \textbf{Resource Efficiency:} Harnessing local resources dramatically curtails the necessity of importing massive construction materials from Earth, leading to logistical and economic efficiencies \cite{Zacny2019}.
    \item \textbf{Quick Setting:} Contrasting with terrestrial concrete that mandates water and extended curing durations, Martian concrete offers rapid setting—a significant advantage in Mars' volatile environment \cite{Lin2016}.
    \item \textbf{Reusability:} The potential to melt and reshape this concrete implies infrastructural adaptability on Mars \cite{Qiao2018}.
\end{itemize}

\paragraph{\textbf{Potential Limitations and Challenges}}
\begin{itemize}
    \item \textbf{Toxicity:} Extended exposure to certain sulfur compounds could pose health risks. As a precaution, Martian habitats might necessitate interior coatings or barriers to guarantee human safety \cite{Kanwal2020}.
    \item \textbf{Thermal Conductivity:} Sulfur concrete's distinct thermal characteristics might influence the insulation capacities of Martian edifices constructed with it \cite{Lin2016}.
    \item \textbf{Tensile Strength:} Analogous to numerous concretes, Martian variants might possess compromised tensile strength, suggesting potential reinforcements for specific applications \cite{Qiao2018}.
\end{itemize}

 Summing up, Martian concrete emerges as a frontrunner for realizing robust habitats and infrastructure on the Red Planet. As we refine our understanding and techniques, this material's prominence in Martian construction endeavors is poised to grow.

\subsubsection{Underground Cities}
 Beyond surface-level habitats, the very substratum of Mars may offer a tantalizing solution to some of the most pressing colonization challenges. Underground cities, or subterranean habitats, can leverage the Martian crust's innate properties to provide shelter, security, and sustainability \cite{Boston2004}.

\paragraph{\textbf{Natural Radiation Shielding}}
\begin{itemize}
    \item \textbf{Protection Depth:} Mars lacks a strong magnetosphere and thick atmosphere, both of which provide Earth with considerable protection from cosmic and solar radiation. By constructing habitats several meters below the surface, colonists can use the regolith and rock as a natural shield, significantly reducing radiation exposure \cite{Dartnell2013}.
    \item \textbf{Consistency:} Unlike makeshift shields that may require maintenance and periodic replacement, the consistent composition of the Martian crust provides enduring protection \cite{Dartnell2013}.
\end{itemize}

\paragraph{\textbf{Thermal Insulation}}
\begin{itemize}
    \item \textbf{Stability:} The Martian surface experiences drastic temperature fluctuations. However, subterranean environments tend to maintain a more stable temperature, reducing the energy required for heating and cooling \cite{Clifford2005}.
    \item \textbf{Heat Retention:} The ground naturally serves as an insulator, helping to trap heat and stabilize interior temperatures, especially during prolonged cold periods \cite{Clifford2005}.
\end{itemize}

\paragraph{\textbf{Resource Extraction}}
\begin{itemize}
    \item \textbf{Water Ice:} Recent discoveries indicate substantial deposits of water ice beneath Mars' surface. Underground habitats can directly access these reserves, providing vital resources for drinking, agriculture, and potentially fuel production \cite{Stuurman2016}.
    \item \textbf{Minerals and Metals:} As we dig deeper into Mars, there is potential to discover and extract valuable minerals and metals that can be utilized for in-situ manufacturing and construction \cite{Wanke1999}.
\end{itemize}

\paragraph{\textbf{Challenges of Subterranean Living}}
\begin{itemize}
    \item \textbf{Excavation:} The act of digging and tunneling on Mars poses significant challenges. Martian regolith's properties, combined with the planet's reduced gravity, necessitate the development of new excavation technologies \cite{Zacny2019}.
    \item \textbf{Air Circulation:} Ensuring a consistent and fresh supply of breathable air within underground habitats demands advanced ventilation systems \cite{Boston2004}.
    \item \textbf{Psychological Aspects:} Extended periods underground could impact colonists' mental well-being. Solutions might include virtual windows, communal gathering areas with simulated natural light, and regular surface excursions \cite{Clancey1999}.
\end{itemize}

 In essence, while underground cities present novel solutions to a myriad of Martian challenges, their realization demands significant advancements in excavation technology, habitat design, and psychological support structures \cite{Boston2001, Cockell2016, Kanas2010}. Yet, as our understanding of Mars deepens, so too does the potential for these subterranean havens to play a pivotal role in human colonization \cite{Zubrin1996, Ferl2012}.

\subsection{Energy Production}
The energy requirements for a Martian colony are multifaceted, requiring a balance between reliability, sustainability, and efficiency \cite{Zubrin1996}. The choice of energy production methods needs to account for Mars's unique environmental conditions. We will assess the potential power yield of different energy sources and analyze their viability.

\subsubsection{Solar Energy}
The solar constant on Mars, \( I_{mars} \), is approximately half of that on Earth, given by Eq. \eqref{eq:solar_constant} \cite{Smith2013}:

\begin{equation}
    I_{mars} \approx 590 \, \text{W/m}^2
    \label{eq:solar_constant}
\end{equation}

Given this, the potential power \( P_{solar} \) from a solar panel of area \( A \) with efficiency \( \eta \) can be calculated as given by Eq. \eqref{eq:solar_power} \cite{Vasavada2007}:

\begin{equation}
    P_{solar} = \eta \times A \times I_{mars}
    \label{eq:solar_power}
\end{equation}

However, due to dust accumulation and reduced daylight hours during Martian winters, the effective efficiency might be reduced by a factor \( f \) as shown in Eq. \eqref{eq:effective_solar_power} \cite{Kinch2015}:

\begin{equation}
    P_{effective} = f \times P_{solar}
    \label{eq:effective_solar_power}
\end{equation}

\subsubsection{Nuclear Energy}
The power \( P_{nuclear} \) produced by a nuclear reactor is given by Eq. \eqref{eq:nuclear_power} \cite{Mason2008}:

\begin{equation}
    P_{nuclear} = \eta_{n} \times m \times E
    \label{eq:nuclear_power}
\end{equation}

where \( \eta_{n} \) is the reactor's efficiency, \( m \) is the mass of the fuel, and \( E \) is the energy density of the fuel. Given the constraints of transporting nuclear fuel from Earth and potential safety concerns, reactors with higher energy densities and efficiencies are preferred \cite{Poston2012}.

\subsubsection{Wind Energy}
The power \( P_{wind} \) that can be harvested from the wind using a turbine is given by Eq. \eqref{eq:wind_power} \cite{Newman2015}:

\begin{equation}
    P_{wind} = \frac{1}{2} \times \rho \times A \times v^3 \times \eta_{w}
    \label{eq:wind_power}
\end{equation}

where \( \rho \) is the atmospheric density of Mars, \( A \) is the area swept by the turbine blades, \( v \) is the wind velocity, and \( \eta_{w} \) is the efficiency of the wind turbine. Given Mars's thin atmosphere, \( \rho \) is significantly lower than on Earth, affecting the viability of wind energy \cite{Murphy2010}.

\subsubsection{Geothermal Energy}
The potential geothermal power \( P_{geo} \) is contingent on the heat flux from Mars's interior as described in Eq. \eqref{eq:geothermal_power} \cite{Plesa2016}:

\begin{equation}
    P_{geo} = \eta_{g} \times A_{g} \times q
    \label{eq:geothermal_power}
\end{equation}

where \( \eta_{g} \) is the geothermal plant's efficiency, \( A_{g} \) is the area from which heat is extracted, and \( q \) is the Martian heat flux. The feasibility of geothermal energy hinges on precise measurements of Mars's internal heat flux and the development of drilling technologies \cite{Grott2012}.

These equations provide a foundational framework for quantifying the potential energy production capabilities of different sources on Mars. Combining these insights with technological advancements and resource availability will inform the energy strategy for Martian colonization \cite{Zubrin1996}.

\subsection{Transportation}

 One of the most formidable challenges in Mars colonization lies in the realm of transportation. Not only does the transportation infrastructure need to be cost-effective, but it must also be reliable and scalable to sustain a growing colony. Solar sail-powered cargo ships offer a promising method that leverages the continuous radiation pressure from the Sun to propel spacecraft \cite{McInnes2004, Johnson2018}.

\subsubsection{Solar Sail-Powered Cargo Ships}

\paragraph{\textbf{Basic Principle}}

Solar sails harness the momentum of photons emitted by the Sun \cite{McInnes2004, Johnson2018}. Though individual photons have no mass, they possess momentum, which can exert a tiny force when absorbed or reflected by a surface. Given a sufficiently large collecting area (the sail), these minute forces sum up to produce tangible propulsion.

The force \( F_{sail} \) exerted on the sail by the photons can be expressed by Eq. \eqref{eq:sail_force}:

\begin{equation}
    F_{sail} = \frac{2 \times I_{solar} \times A_{sail}}{c}
    \label{eq:sail_force}
\end{equation}

where \( I_{solar} \) is the solar irradiance at the sail's location, \( A_{sail} \) is the effective area of the sail, and \( c \) is the speed of light \cite{Johnson2018}.

\paragraph{\textbf{Advantages}}

\begin{itemize}
    \item \textbf{Fuel-free Propulsion:} Once deployed, solar sails require no propellant, thus reducing the mass and complexity of missions \cite{McInnes2004}.
    \item \textbf{Continuous Acceleration:} Unlike chemical rockets which provide thrust in short bursts, solar sails can achieve continuous acceleration as long as they are exposed to sunlight, allowing for potentially higher speeds over long durations \cite{Johnson2018}.
\end{itemize}

\paragraph{\textbf{Challenges}}

\begin{itemize}
    \item \textbf{Initial Acceleration:} Solar sails are not efficient for rapid accelerations; hence, they are best suited for cargo missions or voyages where time is not the most critical factor \cite{McInnes2004}.
    \item \textbf{Sail Material:} The sail must be both highly reflective and exceptionally lightweight. This necessitates the development and deployment of advanced materials \cite{McInnes2004}.
    \item \textbf{Orientation and Control:} Adjusting the sail's angle relative to the Sun allows for navigation. Precision control mechanisms are essential for course adjustments \cite{Johnson2018}.
\end{itemize}

\paragraph{\textbf{Potential Applications for Mars}}

Given the advantages of solar sails, especially for long-duration, propellant-free missions, they could be pivotal for:

\begin{itemize}
    \item \textbf{Cargo Transport:} Delivering equipment, raw materials, or other essentials from Earth or from asteroids to Mars \cite{McInnes2004}.
    \item \textbf{Data Relay:} Establishing a network of solar sail-powered satellites around Mars to facilitate uninterrupted communication with Earth \cite{Johnson2018}.
\end{itemize}

 As our understanding of solar sail dynamics improves, and as material sciences advance, the role of solar sail-powered ships in the interplanetary transportation matrix is set to grow \cite{McInnes2004, Johnson2018}. Their potential for sustainable and cost-effective voyages could redefine the logistics of Martian colonization.

\subsubsection{Rovers and Vehicles for Martian Terrain}

\paragraph{\textbf{Unique Challenges of Martian Terrain}}

Mars presents a range of terrains, from vast plains to towering volcanoes and deep canyons. The following factors significantly influence the design and operation of rovers and vehicles \cite{Murphy2010, Kinch2015}:

\begin{itemize}
    \item \textbf{Low Gravity:} Mars has only about \(0.375\) times Earth's gravity, which affects traction, stability, and the risk of overturning during rapid maneuvers \cite{Plesa2016}.
    
    \item \textbf{Dusty Surface:} Martian regolith consists of fine dust that can clog systems, abrade surfaces, and impact mechanical parts \cite{Kinch2015}.
    
    \item \textbf{Temperature Extremes:} With temperatures often dropping below \(-100^\circ C\), thermal management becomes crucial \cite{Newman2015}.
\end{itemize}

\paragraph{\textbf{Rover Design Principles}}

Considering the Martian challenges, the following design principles are paramount \cite{Yingst2017,Heverly2013}:

\begin{itemize}
    \item \textbf{Robust Mobility:} The rover should feature a multi-wheeled design, often six wheels, to distribute weight and ensure continued mobility even if one wheel malfunctions. The equation governing the force \( F_{traction} \) due to Martian gravity is Eq. \eqref{eq:traction_force}:
    
    \begin{equation}
        F_{traction} = \mu \times m \times g_{mars}
        \label{eq:traction_force}
    \end{equation}
    
    where \( \mu \) is the coefficient of friction, \( m \) is the mass of the rover, and \( g_{mars} \approx 3.71 \, \text{m/s}^2 \) is the acceleration due to gravity on Mars.
    
    \item \textbf{Adaptable Suspension:} To navigate rocky terrains, rovers should have a suspension system that can adjust to irregularities, ensuring all wheels maintain contact with the ground.
    
    \item \textbf{Dust Mitigation:} To counter the abrasive Martian dust, rovers need protective layers on sensitive parts and possibly electrostatic or mechanical dust removal systems.
    
    \item \textbf{Thermal Regulation:} Incorporating heaters and insulators to protect against cold and radiative surfaces to dissipate excessive heat is essential.
\end{itemize}

\paragraph{\textbf{Manned Vehicles}}

More advanced vehicles designed for longer journeys and higher payloads will be required for manned missions and larger colonies \cite{Hoffman2016}. These vehicles would need:

\begin{itemize}
    \item \textbf{Pressurized Cabins:} To protect astronauts from low atmospheric pressure and provide a breathable atmosphere.
    
    \item \textbf{Advanced Navigation Systems:} Given the vast and sometimes featureless Martian landscapes, reliable navigation using onboard systems combined with satellite-based guidance will be crucial.
    
    \item \textbf{Versatility:} Ability to attach trailers, carry tools, and possibly deploy smaller rovers or drones for specialized tasks.
\end{itemize}

 As our understanding of Mars's diverse landscapes expands and as technological advances are made, rovers and vehicles will evolve to address the unique challenges and requirements of the Martian environment, playing a pivotal role in the exploration and colonization of the Red Planet \cite{Squyres2004}.

\subsection{Life Support Systems}

 The sustenance of human life on Mars hinges on developing robust life support systems \cite{Jones2007}. These systems must provide the necessities—air, water, and food—and ensure waste management, temperature regulation, and protection against the Martian environment's inherent challenges. Among these necessities, water stands out as a critical resource for consumption, crop cultivation, and other activities. Given the challenges of transporting large quantities of water from Earth, in-situ water harvesting on Mars becomes an imperative \cite{Eckart2008}.

\subsubsection{Water Harvesting}

\paragraph{\textbf{Significance}}

Water is essential for human survival and many applications in a colony, such as hydroponic agriculture, hygiene, and as a potential source of oxygen through electrolysis. Moreover, water can be split into hydrogen and oxygen, fueling rockets and other energy-intensive processes \cite{Zubrin1996}.

\paragraph{\textbf{Sources of Martian Water}}

While Mars appears arid, it harbors significant amounts of water, chiefly in the form of \cite{Smith2009,Carr2006}:

\begin{itemize}
    \item \textbf{Subsurface Ice:} Large reserves of water ice exist just below the Martian surface, especially at higher latitudes.
    \item \textbf{Atmospheric Moisture:} Though thin, Mars' atmosphere contains water vapor, which can be harvested with suitable technologies.
    \item \textbf{Salt Hydrates:} Certain Martian salts can trap water molecules, a potential, albeit challenging, water source.
\end{itemize}

\paragraph{\textbf{Techniques for Water Harvesting}}

\begin{itemize}
    \item \textbf{Subsurface Ice Mining:} Extracting water from underground ice reserves involves drilling or digging followed by a heating process to convert ice to vapor \cite{Kieffer2006, Hecht2009}. The energy requirement \( E_{extraction} \) for this conversion, considering the latent heat of fusion and vaporization, is given by Eq. \eqref{eq:energy_extraction}:
    
    \begin{equation}
        E_{extraction} = m_{ice} \times (L_{fusion} + L_{vaporization})
        \label{eq:energy_extraction}
    \end{equation}
    
    \item \textbf{Atmospheric Water Extractors:} Devices like adsorption-based harvesters can capture atmospheric water vapor during colder periods and release it during warmer times by heating the adsorbent \cite{Murphy2015}.
    
    \item \textbf{Salt Hydrate Mining:} Extracting water from salt hydrates would involve heating the salts to release the trapped water molecules. This is more energy-intensive and might be viable as a supplementary method \cite{Vaniman2014}.
\end{itemize}

 The methodologies chosen for water harvesting must balance energy costs, efficiency, and the potential environmental impact on Mars \cite{Quinn2007}. Innovations in this domain will lay the foundation for self-sustaining colonies, reducing the reliance on Earth for water resupply missions \cite{Beaty2019}.

\subsubsection{Aeroponics and Food Production}

\paragraph{\textbf{The Need for Sustainable Agriculture on Mars}}

Ensuring a reliable and sustainable food supply is a cornerstone for any long-term human presence on Mars \cite{Wamelink2014}. Traditional soil-based agriculture may be challenging due to the potential presence of harmful perchlorates in Martian soil and the considerable effort required to recreate Earth-like soil conditions \cite{Hecht2009}. This necessitates innovative agricultural methods tailored for Martian conditions.

\paragraph{\textbf{What is Aeroponics?}}

Aeroponics is a method of growing plants without soil, where roots are suspended in air and are periodically misted with nutrient-rich water \cite{Ehrlich2016}. This technique offers several advantages, especially in environments where water and nutrients must be judiciously used, making it an attractive proposition for Mars \cite{Dixon2008}.

\paragraph{\textbf{Advantages of Aeroponics in Martian Context}}

\begin{itemize}
    \item \textbf{Water Efficiency:} Aeroponics can use up to \(90\%\) less water than traditional farming, crucial in a water-scarce environment \cite{Dixon2008}. The water usage \( W \) for a plant in aeroponics is often given by Eq. \eqref{eq:water_usage}:
    
    \begin{equation}
        W = V_{spray} \times f_{cycle} \times t_{growth}
        \label{eq:water_usage}
    \end{equation}
    
    \item \textbf{Rapid Growth:} Plants grow faster due to the continuous availability of oxygen and nutrients directly to their roots \cite{Ehrlich2016}.
    
    \item \textbf{Space Efficiency:} Given the limited space inside habitats, vertical aeroponic farms can maximize the number of crops grown per unit area \cite{Despommier2009}.
    
    \item \textbf{Isolation from Martian Soil:} This method bypasses the challenges of potentially harmful elements in Martian regolith, providing a controlled environment for plant growth \cite{Hecht2009}.
\end{itemize}

\paragraph{\textbf{Challenges and Considerations}}

While aeroponics promises a slew of benefits, it isn't without challenges \cite{Stoner1996, Dreschel1988}:

\begin{itemize}
    \item \textbf{System Failures:} The reliance on misting systems means mechanical or electrical failures could jeopardize an entire crop.
    
    \item \textbf{Nutrient Delivery:} Ensuring a consistent and optimal mix of nutrients in the misting solution is crucial for plant health and yield \cite{Ehrlich2016}.
    
    \item \textbf{Microgravity Concerns:} While Mars has \(0.375\) times Earth's gravity, the effects of this reduced gravity on aeroponic systems and plant growth are yet to be fully understood \cite{Wheeler2008}.
\end{itemize}

 As technology progresses and our understanding of Martian conditions improves, integrating aeroponics into Martian colonies' food production strategies will be a vital step toward achieving self-sustainability \cite{Wheeler2017}.

\subsubsection{Algae Bioreactors}

\paragraph{\textbf{The Role of Algae in Space Colonization}}

In the context of space colonization, algae are incredibly versatile microorganisms \cite{Eichler2017}. They can potentially address some of the most pressing challenges of establishing a self-sustaining human presence on Mars: oxygen production, waste recycling, and food production \cite{Niederwieser2018}. Algae bioreactors are systems that cultivate these microorganisms under controlled conditions, optimizing their growth and metabolic outputs \cite{Merzlyak2005}.

\paragraph{\textbf{Benefits of Algae Bioreactors in a Martian Environment}}

\begin{itemize}
    \item \textbf{Oxygen Production:} algae consume carbon dioxide through photosynthesis and release oxygen. This process can be modeled as Eq. \eqref{eq:photosynthesis}:
    
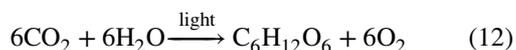
    \begin{equation}
        6 \text{CO}_2 + 6 \text{H}_2\text{O} \xrightarrow{\text{light}} \text{C}_6\text{H}_{12}\text{O}_6 + 6 \text{O}_2
        \label{eq:photosynthesis}
    \end{equation}
    
    Here, the conversion efficiency and the rate at which oxygen is produced depend on light availability, algae species, and nutrient conditions \cite{Blanken2017}.
    
    \item \textbf{Nutrient Recycling:} Algae can process wastewater, absorbing nutrients like nitrogen and phosphorus, thereby purifying it \cite{Bilanovic2009}.
    
    \item \textbf{Food Source:} Certain algae strains, like Spirulina, are rich in proteins and essential nutrients, making them a valuable supplemental food source \cite{Vonshak1997}.

    \item \textbf{Biofuel Production:} Algae can also be cultivated for lipid production, which can be converted into biofuels, serving as a potential energy resource \cite{Mata2009}.
\end{itemize}

\paragraph{\textbf{Designing Bioreactors for Mars}}

While the concept of algae bioreactors is not new, designing them for Martian conditions presents unique challenges:

\begin{itemize}
    \item \textbf{Temperature Regulation:} Mars' extreme temperature fluctuations necessitate systems insulating algae cultures and maintaining a stable temperature \cite{Schuerger2004}.
    
    \item \textbf{Radiation Protection:} Algae cultures must be shielded from harmful cosmic and solar radiation \cite{Verseux2016}.
    
    \item \textbf{Optimal Light Source:} Given the greater distance from the Sun and frequent dust storms, ensuring a consistent and optimal light source—potentially using LEDs—is essential for maximizing photosynthesis \cite{Massa2017}.
\end{itemize}

\paragraph{\textbf{Future Prospects}}

Integrating algae bioreactors into Martian habitats could be a game-changer, streamlining multiple life support processes \cite{Hezard2019}. Research into hardier strains of algae optimized for Martian conditions and continuous improvements in bioreactor design will be essential in the journey to make Mars a second home for humanity.

\subsubsection{Magnetic Shield Deployment}

\paragraph{\textbf{The Radiation Threat on Mars}}

One of the significant challenges of prolonged human habitation on Mars is the pervasive cosmic and solar radiation. Unlike Earth, which has a robust magnetic field and thick atmosphere that deflects and attenuates much of this radiation, Mars lacks a global magnetic field and has a thin atmosphere, making its surface more exposed to harmful radiation \cite{Schuerger2004, Verseux2016}. These radiations pose severe health risks, including increased chances of cancer, tissue damage, and other long-term health issues.

\paragraph{\textbf{Magnetic Shielding: A Potential Solution}}

Magnetic shields serve as active radiation protection systems by creating a magnetic field strong enough to deflect charged particles, mimicking, to an extent, Earth's natural protective magnetosphere. The concept is rooted in magnetohydrodynamics: charged particles, like those in cosmic and solar radiation, will follow magnetic field lines and can be diverted away from protected areas \cite{Verseux2016, Schuerger2004}.

\paragraph{\textbf{Working Principle}}

The underlying principle can be encapsulated in the Lorentz force law (Eq. \eqref{eq:lorentz}:

\begin{equation}
    \mathbf{F} = q(\mathbf{v} \times \mathbf{B})
    \label{eq:lorentz}
\end{equation}

where \( \mathbf{F} \) is the force experienced by the charged particle, \( q \) is its charge, \( \mathbf{v} \) is its velocity, and \( \mathbf{B} \) is the magnetic field \cite{Schuerger2004}.

In the context of the shield, charged particles from cosmic and solar radiation would experience a force due to the artificial magnetic field, changing their trajectories and preventing them from reaching the protected area.

\paragraph{\textbf{Implementation Challenges}}

\begin{itemize}
    \item \textbf{Power Requirements:} Generating a magnetic field strong enough to offer significant protection would require substantial energy. The power \( P \) required can be related to the magnetic field strength \( B \) and the volume \( V \) of the shielded area through certain constants and efficiency factors (Eq. \eqref{eq:power}):
    
    \begin{equation}
        P \propto \eta B^2 V
        \label{eq:power}
    \end{equation}
    where \( \eta \) represents efficiency and proportionality factors \cite{Massa2017}.

    \item \textbf{Magnetic Field Configuration:} Uniform magnetic fields might not be the most effective; more complex field configurations, like toroidal or helical fields, might offer better protection.

    \item \textbf{Infrastructure and Weight:} The infrastructure to generate and maintain these magnetic fields, especially on a large scale, would be massive and pose transportation and deployment challenges \cite{Verseux2016}.
\end{itemize}

\paragraph{\textbf{Future Outlook}}

Despite the challenges, the deployment of magnetic shields on Mars holds promise. With advancements in superconductor technology, energy storage, and magnet design, magnetic shielding could become a pivotal technology to safeguard future Martian settlers against relentless radiation \cite{Hezard2019, Verseux2016}.

\subsection{Communication: Mars-Earth Relay Systems}

Maintaining robust, reliable, and real-time communication between Mars and Earth is paramount for the success of a Martian colony. The vast distance and the potential for direct signal obstruction by the Sun or other celestial bodies mandate a sophisticated relay system. Utilizing special stable points in space, known as Lagrange points, particularly the Mars L2 point, offers a potential solution \cite{Mata2009, Bilanovic2009}.

\subsubsection{The Mars L2 Lagrange Point}
Lagrange points are governed by the gravitational forces of two primary bodies and the orbital motion of the smaller body \cite{Murray1999}. For a satellite to remain at a Lagrange point, the effective gravitational potential at that point should be at a local maximum or minimum \cite{Bate1971}. The effective potential \( V \) is given by Eq. \eqref{eq:potential}:

\begin{equation}
V = \frac{1}{2} \omega^2 r^2 + \frac{G M_{sun}}{r} + \frac{G M_{mars}}{|r - R|}
\label{eq:potential}
\end{equation}

Where:
\begin{itemize}
    \item \( \omega \) is Mars' angular speed about the Sun.
    \item \( r \) is the distance from the Sun to the satellite.
    \item \( R \) is the distance from the Sun to Mars.
    \item \( G \) is the gravitational constant.
    \item \( M_{sun} \) and \( M_{mars} \) are the masses of the Sun and Mars, respectively.
\end{itemize}
For the L2 point, the effective potential \( V \) should have a local maximum.

\subsubsection{Advantages of Using the Mars L2 Point}
\begin{itemize}
    \item \textbf{Constant Line of Sight:} Satellites at the Mars L2 point will always have a clear line of sight to Mars and Earth (when Mars is not positioned directly between the Sun and Earth), ensuring uninterrupted communication \cite{Hand2010}.
    \item \textbf{Reduced Communication Latency:} Having a dedicated relay point can help achieve faster communication speeds than a scenario without relay satellites.
    \item \textbf{Scalability:} The L2 point can host multiple satellites, creating a network for redundancy and increased bandwidth.
\end{itemize}

\subsubsection{Implementation Challenges}
Besides the mentioned challenges, placing a satellite at L2 requires careful calculating of its required velocity to maintain its position at the Lagrange point. The required centripetal force \( F_c \) for the satellite to orbit the Sun in sync with Mars is Eq. \eqref{eq:centripetal}:

\begin{equation}
F_c = m \omega^2 (R + \delta)
\label{eq:centripetal}
\end{equation}

Where:
\begin{itemize}
    \item \( m \) is the mass of the satellite.
    \item \( \delta \) is the distance from Mars to the L2 point.
\end{itemize}
This centripetal force should be balanced by the net gravitational force from the Sun and Mars at the L2 point. Challenges include:
\begin{itemize}
    \item \textbf{Orbital Stability:} While the Mars L2 point is a point of equilibrium, it is not entirely stable. Satellites would require periodic propulsion-based adjustments to remain in position \cite{Howell2010}.
    \item \textbf{Distance from Mars:} The L2 point is considerably farther from Mars than a typical Martian orbit, meaning satellites placed here would need powerful transmission systems to relay signals effectively to and from the Martian surface.
    \item \textbf{Protection:} Satellites in this position are exposed to higher solar radiation levels, requiring enhanced protective measures to ensure longevity.
\end{itemize}

The Mars-Earth communication challenge is a technical hurdle and a lifeline for Martian settlers \cite{Wall2015}. The Mars L2 Lagrange point presents a promising solution to this challenge. As technology advances, this strategy and other communication modalities could pave the way for seamless interplanetary communication, bridging the gap between the two worlds.

\subsection{Space and Surface Mobility}
Space and surface mobility solutions are fundamental to the success of any Martian colonization effort \cite{Landis2011}. These solutions should address the Martian environment's unique challenges while ensuring safety, efficiency, and reliability. This subsection details the advancements in spacesuits, rovers, and other mobility solutions tailored for Mars.

\subsubsection{Advanced Spacesuits}

\textbf{Design Requirements:}
Martian spacesuits must satisfy numerous criteria:

\begin{itemize}
    \item \textbf{Radiation Protection:} The thin Martian atmosphere means that spacesuits must shield astronauts from cosmic and solar radiation \cite{Hassler2014, Zeitlin2013}.
    \item \textbf{Thermal Regulation:} With extreme temperature variations on Mars, spacesuits must provide insulation and cooling \cite{Kosmo1993, Carr2003}.
    \item \textbf{Dust Protection:} Martian dust is fine and pervasive. Spacesuits must prevent this dust from infiltrating and compromising the suit's systems or the astronaut's health \cite{Thomas2005}.
\end{itemize}

\textbf{Technological Advancements:}
\begin{itemize}
    \item \textbf{Layered Shielding:} Using layered composite materials with different properties to enhance radiation protection \cite{Townsend2001}. The effectiveness, \( E \), of a shield can be described as Eq. \eqref{eq:shielding}:    
    \begin{equation}
    E = \int_{0}^{d} \mu(x) \, dx
    \label{eq:shielding}
    \end{equation}
    
    where \( d \) is the thickness of the shield and \( \mu(x) \) is the attenuation coefficient of the material at position \( x \) \cite{Tripathi2006}.
    \item \textbf{Regenerative Cooling Systems:} Harnessing phase change materials and endothermic reactions to absorb excess heat, helping in temperature regulation \cite{Kosmo1993}.
    \item \textbf{Electrostatic Dust Repellents:} Incorporating materials that can hold a static charge, repelling the Martian dust \cite{Melchiorri2001}.
\end{itemize}

\subsubsection{Rovers and Surface Vehicles}

\textbf{Design Considerations:}
\begin{itemize}
    \item \textbf{Energy Efficiency:} Given the extended mission durations and the remoteness of operations, Martian rovers must optimize energy consumption \cite{Arvidson2011, Squyres2004}.
    \item \textbf{Traction and Stability:} The Martian surface is a mix of dunes, rocks, and ravines. Rovers need to navigate this terrain reliably \cite{Mishkin2004}.
\end{itemize}

\textbf{Technological Innovations:}
\begin{itemize}
    \item \textbf{Hybrid Propulsion Systems:} Combining solar and nuclear energy sources to ensure consistent energy supply \cite{Zubrin1996}.
    \item \textbf{Adaptive Wheel Technology:} Wheels that can change shape or stiffness depending on the terrain, optimizing traction and reducing wear \cite{Trease2014}.
    \item \textbf{Autonomous Navigation:} Incorporating machine learning algorithms to allow rovers to adapt to new terrains and obstacles without constant input from mission control \cite{Maimone2007}. The pathfinding algorithm, \( P \), can optimize routes based on terrain difficulty, \( T \), and energy requirements, \( E \) (Eq. \eqref{eq:pathfinding}):
    
    \begin{equation}
    P = \min \left( \alpha T + \beta E \right)
    \label{eq:pathfinding}
    \end{equation}
    
    where \( \alpha \) and \( \beta \) are weighting coefficients.
\end{itemize}

\subsubsection{Other Mobility Solutions}
\begin{itemize}
    \item \textbf{Portable Habitats:} Compact habitats that can be deployed during extended missions away from the primary base. These habitats would use inflatables and rigid sections, balancing portability and protection \cite{Horneck2006, Chamberlain2007}.
    \item \textbf{Drones:} Martian drones, powered by thin-film solar panels, could scout areas, create topographical maps, or even transport small payloads \cite{Anderson2019}. Their flight dynamics must account for the thin Martian atmosphere, using larger wing surfaces or more efficient propulsion methods \cite{Bousquet2018}.
\end{itemize}

Advanced mobility solutions are desirable and imperative for a successful Martian colonization. They enable exploration, ensure safety, and enhance astronauts' overall quality of life \cite{Lavery2010, Crawford2015}. As technology progresses, these solutions will undoubtedly evolve, becoming even more tailored to the unique challenges of the Red Planet.

\section{Human Factors}

Human factors address the intersection of humans and the environments in which they operate. Establishing a Martian colony is not solely a technological challenge; it's deeply rooted in the human experience \cite{Kanas2013}. Adapting to a new world while ensuring physical, psychological, and social well-being will be crucial to any off-world endeavor's long-term success and sustainability \cite{Sandberg2013}.

\subsection{Psychological Challenges of Isolation and Potential Solutions}

Life on Mars presents unprecedented challenges of isolation, confinement, and distance from Earth \cite{Kanas2008}. The psychological stressors of living in such conditions—far removed from the familiar surroundings of our home planet—can lead to a myriad of issues, from depression and anxiety to interpersonal conflicts \cite{Ritsher2007}. Addressing these is paramount for the well-being of astronauts and the success of Martian missions.

\subsubsection{Sources of Isolation and Confinement Stressors}
\begin{itemize}
    \item \textbf{Distance from Earth:} Communication delays, the physical divide, and the understanding that returning home isn't immediate can cause feelings of isolation \cite{Kanas2008}.
    \item \textbf{Limited Social Interactions:} Being confined with the same group for extended periods can lead to monotony and interpersonal tensions \cite{Bishop2000}.
    \item \textbf{Harsh External Environment:} The inability to step outside freely without a spacesuit and the stark Martian landscape can amplify feelings of confinement \cite{Harrison2005}.
\end{itemize}

\subsubsection{Potential Psychological Impacts}
These stressors can lead to:
\begin{itemize}
    \item Decreased mood and morale \cite{Palinkas2003, Slack2004}.
    \item Reduced cognitive performance \cite{Slack2004}.
    \item Sleep disturbances \cite{Palinkas2003}.
    \item Interpersonal conflicts and breakdowns in team cohesion \cite{Palinkas2003, Slack2004}.
\end{itemize}

\subsubsection{Proposed Solutions}
\begin{itemize}
    \item \textbf{Advanced Training:} Pre-mission training to develop resilience, stress management, and conflict resolution skills \cite{Palinkas2003}.
    \item \textbf{Virtual Reality (VR) Environments:} Using VR to simulate varied and familiar environments, offering a respite from the monotony of Mars \cite{Slack2004}.
    \item \textbf{Telemedicine and Psychological Support:} Providing access to psychological professionals on Earth, catering to individual needs and delivering interventions when necessary \cite{Palinkas2003}.
    \item \textbf{Habitat Design:} Creating living spaces that inhabitants can modify, introducing elements of Earth's nature, and ensuring private spaces for personal time.
    \item \textbf{Engaging Activities:} Organizing structured recreational and creative activities to combat monotony and boost morale.
    \item \textbf{Communication Technology:} While real-time communication is challenging due to the distance, technologies that can bridge this gap or minimize its effect can be essential.
\end{itemize}

The isolation and confinement challenges posed by Martian life are significant, but they can be mitigated with a combination of training, technology, and innovative solutions \cite{Palinkas2003}. Addressing the psychological well-being of astronauts is as crucial as any technological solution, ensuring the long-term success of Martian colonization efforts \cite{Slack2004}.

\subsection{Training for Self-Sufficiency Given the Communication Delay with Earth}

The vast distance between Mars and Earth results in a substantial communication delay, ranging from 4 to 24 minutes one way \cite{Mishkin2004}. This delay, combined with the potential for communication blackouts, means Martian colonists cannot rely on real-time support from Earth for many tasks. As such, training for self-sufficiency becomes paramount for ensuring the colony's safety, well-being, and productivity.

\subsubsection{Challenges Posed by Communication Delays}
\begin{itemize}
    \item \textbf{Emergency Response:} Immediate decision-making is required in emergencies, with no time to consult Earth.
    \item \textbf{Routine Operations:} Day-to-day activities, especially those involving complex machinery or systems, can be hindered by waiting for guidance \cite{Mishkin2004}.
    \item \textbf{Medical Situations:} Medical emergencies require swift interventions, and specialized knowledge might be necessary beyond the crew's expertise.
\end{itemize}

\subsubsection{Training Approaches for Self-Sufficiency}
\begin{itemize}
    \item \textbf{Scenario-Based Training:} Trainees face many simulated challenges, emergencies, and day-to-day activities, emphasizing decision-making without Earth's input. The efficacy, \( E_t \), of this training, can be gauged by the ratio of successful resolutions to challenges presented (Eq. \eqref{eq:training_efficacy}):

    \begin{equation}
    E_t = \frac{\text{number of successful resolutions}}{\text{number of presented challenges}}
    \label{eq:training_efficacy}
    \end{equation}

    \item \textbf{Cross-Training:} Members are trained in multiple disciplines, ensuring redundancy in skill sets. For instance, engineers might learn basic medical procedures, while medics might learn about habitat maintenance.

    \item \textbf{Utilizing AI and Advanced Simulations:} Embed AI tools and simulations in Martian systems to guide colonists through complex tasks, offering the next best thing to real-time human expertise.

    \item \textbf{Continuous Learning Modules:} Given Earth's mission duration and technological advancements, colonists should have access to learning modules that update them with new techniques, research, and solutions.

    \item \textbf{Behavioral and Psychological Training:} training should also focus on conflict resolution, stress management, and leadership in isolation to ensure team cohesion and mental well-being \cite{Palinkas2003}.
\end{itemize}

\subsubsection{Feedback and Iterative Improvement}
Post-scenario debriefings should be a crucial component of the training process. Analyzing outcomes, discussing alternative solutions, and incorporating lessons learned ensures iterative improvement. This feedback loop, denoted as \( F \), can be defined by the change in efficacy over training scenarios (Eq. \eqref{eq:feedback_loop}):

\begin{equation}
F = \frac{\Delta E_t}{\text{number of training scenarios}}
\label{eq:feedback_loop}
\end{equation}

The communication delay with Earth is more than a technological challenge; it tests the Martian colonists' adaptability, resilience, and autonomy \cite{Mishkin2004}. Comprehensive training that prioritizes self-sufficiency, combined with the tools and knowledge to act independently, will be foundational to the success and survival of a Martian colony.

\subsection{Health Considerations in Reduced Gravity}

Mars has approximately 38\% of Earth's gravity, which presents unique challenges for human health and physiology \cite{Clément2005, Bloomberg2015}. Prolonged exposure to reduced gravity can cause several physiological changes, some of which can impact the health and performance of Martian settlers \cite{Clément2005, Williams2009}. Addressing and mitigating these effects is crucial to ensure the well-being and functionality of humans on Mars.

\subsubsection{Bone Density Loss}
\begin{itemize}
    \item \textbf{Description:} In a low-gravity environment, bones experience less mechanical stress, decreasing bone mineral density. This can result in osteopenia or osteoporosis, conditions characterized by weak and brittle bones \cite{LeBlanc2000, Smith2012}.

    \item \textbf{Quantifying Loss:} The rate of bone density loss, \( R_b \), can be approximated as Eq. \eqref{eq:bone_loss_rate}:

    \begin{equation}
    R_b = -k \times (g_m - g_e)
    \label{eq:bone_loss_rate}
    \end{equation}
    
    Where \( k \) is a proportionality constant, \( g_m \) is the gravity on Mars (3.71 m/s\(^2\)), and \( g_e \) is the gravity on Earth (9.81 m/s\(^2\)) \cite{Lang2011}.

    \item \textbf{Mitigation Strategies:} Weight-bearing exercises, medication, and dietary supplements rich in calcium and vitamin D \cite{Smith2012}.
\end{itemize}

\subsubsection{Muscle Atrophy}
\begin{itemize}
    \item \textbf{Description:} Reduced gravity reduces muscle load, leading to muscle weakening or atrophy over time \cite{LeBlanc2000, Trappe2009}.

    \item \textbf{Quantifying Atrophy:} Muscle atrophy rate, \( R_m \), can be represented similarly to bone density loss (Eq. \eqref{eq:muscle_loss_rate}):

    \begin{equation}
    R_m = -k' \times (g_m - g_e)
    \label{eq:muscle_loss_rate}
    \end{equation}
    
    Where \( k' \) is another proportionality constant \cite{Trappe2009}.

    \item \textbf{Mitigation Strategies:} Regular resistance training, electrostimulation, and protein-rich diets \cite{Fitts2010}.
\end{itemize}

\subsubsection{Cardiovascular Changes}
\begin{itemize}
    \item \textbf{Description:} Reduced gravity can lead to orthostatic intolerance, fluid redistribution, and potential changes in blood volume \cite{Convertino2002, Hargens2013}.

    \item \textbf{Mitigation Strategies:} Compression garments, cardiovascular exercises, and hydration management \cite{Convertino2002}.
\end{itemize}

\subsubsection{Visual Impairment}
\begin{itemize}
    \item \textbf{Description:} Changes in intracranial pressure due to fluid shifts in reduced gravity can affect vision \cite{Mader2011}.

    \item \textbf{Mitigation Strategies:} Monitoring eye health regularly, therapeutic eyewear, and adjusting ambient cabin pressure \cite{Mader2011, Law2017}.
\end{itemize}

Living in Mars' reduced gravity will undeniably impact human health. However, comprehensive understanding, proactive monitoring, and effective countermeasures can minimize the adverse effects \cite{Clément2005}. The success of Martian colonization depends not just on technological innovations but also on safeguarding the health of its inhabitants in this unique environment \cite{Bloomberg2015}.

\section{Economic Considerations}

\subsection{Cost Analysis of Using In-Situ Resources vs. Transport from Earth}

The economics of Martian colonization is a crucial aspect that influences the feasibility and sustainability of the mission \cite{Zubrin1996, Foust2017}. One primary consideration is whether to use in-situ Mars resources or transport necessary materials and goods from Earth \cite{Crawford2004}. This subsection outlines a cost-benefit analysis of these two approaches, considering various factors like transportation costs, technological requirements, and resource availability \cite{Larson2003}.

\subsubsection{Transport Costs from Earth}
\begin{itemize}
    \item \textbf{Description:} Launching payloads from Earth incurs considerable costs. These costs encompass the rocket's price and infrastructure, labor, and additional resources to ensure a successful launch \cite{Foust2017, Sercel2019}.

    \item \textbf{Quantifying Transport Costs:} Let the cost of transporting a unit mass, \( C_t \), from Earth to Mars be defined as Eq. \eqref{eq:transport_cost}:

    \begin{equation}
    C_t = C_{launch} + C_{infra} + C_{labor} + C_{misc}
    \label{eq:transport_cost}
    \end{equation}

    Where:
    \begin{itemize}
        \item \( C_{launch} \) is the cost of the rocket and fuel \cite{Foust2017}.
        \item \( C_{infra} \) is the cost of infrastructure.
        \item \( C_{labor} \) is the cost associated with manpower.
        \item \( C_{misc} \) includes other miscellaneous costs.
    \end{itemize}
\end{itemize}

\subsubsection{In-Situ Resource Utilization (ISRU) Costs}
\begin{itemize}
    \item \textbf{Description:} Using Martian resources, like water and regolith, to produce necessities such as fuel, breathable air, and building materials can potentially reduce the need for transportation from Earth \cite{Rapp2008, Sanders2015}.

    \item \textbf{Quantifying ISRU Costs:} The cost of producing a unit mass using ISRU, \( C_{isru} \), can be formulated as Eq. \eqref{eq:isru_cost}:

    \begin{equation}
    C_{isru} = C_{tech} + C_{op} + C_{maint}
    \label{eq:isru_cost}
    \end{equation}

    Where:
    \begin{itemize}
        \item \( C_{tech} \) is the cost of establishing the ISRU technology on Mars \cite{Rapp2008}.
        \item \( C_{op} \) represents the operational costs, including energy.
        \item \( C_{maint} \) is the maintenance and potential repair costs.
    \end{itemize}
\end{itemize}

\subsubsection{Cost-Benefit Analysis}
Comparing \( C_t \) and \( C_{isru} \) can provide insights into the economic viability of each approach \cite{Duke2003}. However, it's also vital to factor in non-monetary costs like the time delay of transporting resources, the risk associated with launches, and the potential for technological failures or inefficiencies in ISRU processes \cite{Zubrin1996}.

A comprehensive economic model integrating direct and indirect costs, potential risks, and uncertainties will be pivotal in decision-making \cite{Duke2003}. As ISRU technology matures and scales up, it could become a more economically viable solution than interplanetary transport's significant costs and challenges \cite{Sanders2015}.

\subsection{Potential Economic Incentives for Private Companies to Invest in Mars Colonization}

The challenge of Mars colonization is not just a scientific and engineering endeavor; it's also an economic one. The scale and complexity of colonization require significant capital, making private-sector investment indispensable \cite{Zubrin1996, Foust2017}. Several incentives could attract private companies to commit funds and resources to Mars exploration and settlement.

\subsubsection{Extraction and Utilization of Martian Resources}
\begin{itemize}
    \item \textbf{Description:} Mars hosts a variety of resources, including water-ice, metals, and rare minerals \cite{Beaty2015}. The potential to extract and utilize these resources for both Martian and Earth-based applications is vast.
  
    \item \textbf{Economic Model:} Let the expected profit, \( P_{res} \), from resource extraction be defined as Eq. \eqref{eq:resource_profit}:

    \begin{equation}
    P_{res} = R_{sell} - (C_{extraction} + C_{transport})
    \label{eq:resource_profit}
    \end{equation}

    Where:
    \begin{itemize}
        \item \( R_{sell} \) is the revenue from selling the extracted resources.
        \item \( C_{extraction} \) is the cost of extraction on Mars.
        \item \( C_{transport} \) is the cost of transporting the resources, if applicable.
    \end{itemize}
\end{itemize}

\subsubsection{Technological Development and Intellectual Property}
\begin{itemize}
    \item \textbf{Description:} The challenges of Mars colonization will necessitate the development of novel technologies, which can have applications on Earth or be licensed to other entities \cite{Vertesy2017}.

    \item \textbf{Economic Model:} The potential revenue, \( R_{tech} \), from technological development can be defined by Eq. \eqref{eq:tech_revenue}:

    \begin{equation}
    R_{tech} = R_{license} + R_{adapt}
    \label{eq:tech_revenue}
    \end{equation}

    Where:
    \begin{itemize}
        \item \( R_{license} \) is the revenue from licensing the technology.
        \item \( R_{adapt} \) is the revenue from adapting the technology for Earth-based or other space-based applications.
    \end{itemize}
\end{itemize}

\subsubsection{Tourism and Real Estate Development}
\begin{itemize}
    \item \textbf{Description:} As the viability of Mars as a habitat increases, there's potential for tourism and even real estate ventures on the red planet \cite{Foust2017, Launius2018}.

    \item \textbf{Economic Model:} Projected revenue, \( R_{tourism} \), can be modeled as Eq. \eqref{eq:tourism_revenue}:

    \begin{equation}
    R_{tourism} = N_{visitors} \times (T_{ticket} + T_{stay})
    \label{eq:tourism_revenue}
    \end{equation}

    Where:
    \begin{itemize}
        \item \( N_{visitors} \) is the number of tourists.
        \item \( T_{ticket} \) is the cost of a ticket to Mars.
        \item \( T_{stay} \) is the expenditure on Mars for stay, experiences, etc.
    \end{itemize}
\end{itemize}

\subsubsection{Collaborations and Public Relations}
\begin{itemize}
    \item \textbf{Description:} Being part of Mars colonization projects can significantly boost a company's public image, leading to collaborations, partnerships, and enhanced brand value \cite{Gerardi2019}.
    
    \item \textbf{Economic Impact:} Although intangible and hard to quantify, such collaborations can open doors to government contracts, research partnerships, and increased market share due to enhanced brand perception.
\end{itemize}

\section{Recommendations for Future Research}

While this publication provides a comprehensive framework for Mars colonization, numerous areas necessitate further research to refine and optimize the proposed solutions. Here are some pivotal points for future inquiry:

\begin{itemize}
    \item \textbf{Resource Quantification:} A more detailed mapping of Martian resources, such as subsurface water-ice deposits and metal concentrations, will be essential to guide in-situ resource utilization strategies.
    
    \item \textbf{Technological Refinement:} Technologies such as aeroponics, algae bioreactors, and magnetic shield deployment can benefit from iterative prototyping and testing in Martian analog environments on Earth.
    
    \item \textbf{Economic Models:} Advanced economic modeling, incorporating uncertain variables like future technological advancements, potential policy changes, and evolving market dynamics, is necessary for long-term mission planning.
    
    \item \textbf{Human Health:} Continued research into the physiological effects of reduced gravity on human health, especially on long-duration missions, is pivotal. Developing effective countermeasures against radiation and potential psychological challenges must also be prioritized.
    
    \item \textbf{Infrastructure Durability:} The long-term resilience of Martian habitats, especially against Martian dust storms, temperature variations, and radiation, requires further study. Feedback from such studies can inform design improvements.
    
    \item \textbf{Communication Systems:} While using the Mars L2 Lagrange point for a communication relay offers promise, such a system's efficiency, redundancy, and longevity need rigorous testing and validation.
\end{itemize}

By addressing these research points, future iterations of Mars colonization blueprints can achieve greater accuracy, efficiency, and safety, ensuring humanity's next step into the cosmos is confident and informed.

Mars colonization presents a frontier not just for scientific exploration but also for economic opportunities. Forward-thinking companies can leverage these potential incentives to pave the way for humanity's next giant leap while ensuring profitable returns on their investments.

\section{Conclusion}

As the contours of space exploration in the 21st century become sharper focus, Mars emerges not as a distant, aspirational target but as a tangible frontier for human colonization. The challenge of establishing a sustainable presence on the Red Planet is marked by intricacies that extend beyond mere engineering feats, encompassing critical environmental, technological, physiological, and economic dimensions. Our discourse underscores several key imperatives. First, the strategic utilization of in-situ Martian resources is pivotal for mission sustainability and the economic feasibility of prolonged colonization efforts. Second, the human component — the astronauts — demands rigorous attention, with their physiological and psychological health standing as a central pillar for mission success. Third, while technology provides vital tools in this endeavor, its application must be discerning and tailored to Mars's unique challenges and opportunities. The economic architecture of Mars colonization should be grounded in pragmatism, ensuring that investments yield tangible returns, both in terms of scientific gains and potential future economic benefits. Ultimately, the Martian colonization effort is a testament to humanity's perpetual drive for exploration and understanding. As we chart this new chapter in our cosmic journey, our strategies must be informed by a synthesis of caution, innovation, and unwavering commitment to scientific rigor.










\printcredits


\bibliographystyle{cas-model2-names}

\sloppy
\bibliography{cas-refs}



\end{document}